# A DHT Based Measure of Randomness

Renuka Kandregula

*Abstract:* This paper presents a new discrete Hilbert transform (DHT) based measure of randomness for discrete sequences. The measure has been used to test three different classes of sequences with satisfactory results.

## Introduction

Random numbers are used in different fields as in cryptography for generating encryption keys, in simulating and modeling complex phenomena and for selecting random samples from larger data sets. Random numbers may be classified as pseudo-random and true-random. True-random numbers are unpredictable and cannot be generated by a computer algorithm. True random numbers can only be generated by physical processes. Randomness into computers is introduced in the form of pseudo-random numbers. Pseudo-random numbers are not truly random.

Terry Ritter has provided a survey of different randomness tests [1]. For measure of randomness, one may use the computational complexity idea of Kolmogorov [2], or various transform based measures (e.g. [3]) as well as other tests [4].

Here we propose a new measure of randomness based on the discrete Hilbert transform [5]-[9]. We will substantiate this measure by testing it on prime-reciprocal [10]-[15] as well as sequences obtained by the use of random shifts and the use of random numbers generated by the computer [16]-[17].

## Prime Reciprocal Sequences

Prime reciprocal or D sequences are obtained in expansions of fractions or irrational numbers and thus are "decimal" sequences to arbitrary bases [10]-[15]. The basic method of the generating the binary d-sequences is given in [12].

For a certain class of decimal sequences of l/p, p prime, the d-sequence is given by:

$$a_i = \left(2^i \bmod p\right) \bmod 2$$

For example the representation of 1/19 in a base 2 d-sequence using the above notation would be: 000011010111100101

# Random Switch Pseudo-Random Sequences

Switch sequences are pseudo-random numbers which are generated by switching the bits in a binary sequence of a finite length. These switches are chosen at random.

For example let us consider a finite length binary sequence of 100 having fifty 0's and fifty 1's. This will look like Figure 1 after plotting it.

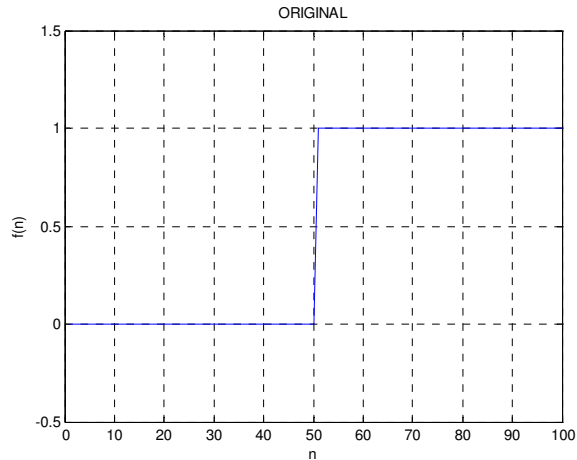

Figure 1. A sequence of 50 0s and 50 1s

To obtain the shift register sequence, we switch two bits random from the first half and the second half of the sequence. 0's are switched to 1's and 1's are switched to 0's.

Switching the 5th, 36th, 56th and the 70th bits, we obtain the two switch sequence as below:

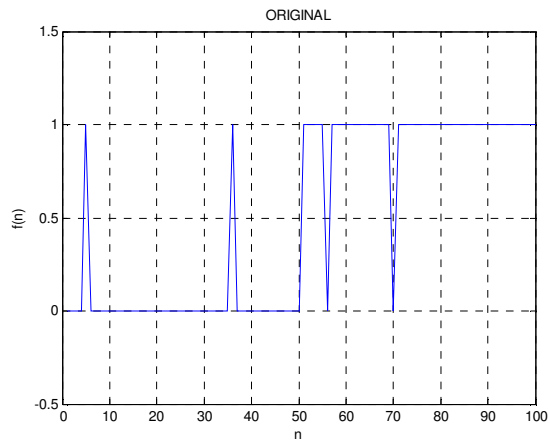

Figure 2. Random sequence derived from previous example

The idea behind generating random sequences in this manner is that of complexity. Each such switch implies higher complexity in the description of the sequence. But certainly, the fact that we use random switches means that it cannot be used as an efficient generator of random sequences. It can, however, serve as a candidate random sequence on which our proposed randomness measure will be tested.

## The Basic Discrete Hilbert Transform

The basic Discrete Hilbert Transform (DHT) of discrete data f(n)  where n = (-∞,…,-1,0,1,…,∞) was given by Kak [5]:

$$DHT\{f(n)\} = g(k) = \begin{cases} \dfrac{2}{\pi} \sum_{n\ odd} \dfrac{f(n)}{k-n}; & k\ even \\ \dfrac{2}{\pi} \sum_{n\ even} \dfrac{f(n)}{k-n}; & k\ odd \end{cases} \tag{1}$$

The inverse Discrete Hilbert Transform (DHT) is given as:

$$f(n) = \begin{cases} -\dfrac{2}{\pi} \sum_{k\ odd} \dfrac{g(k)}{n-k}; & n\ even \\ -\dfrac{2}{\pi} \sum_{k\ even} \dfrac{g(k)}{n-k}; & n\ odd \end{cases} \tag{2}$$

The Hilbert transform has many applications in signal processing, imaging, modulation and demodulation, determination of instantaneous frequency and in cryptography.

The discrete Hilbert transform (DHT) has several forms [6]-[8]. In [9], an application of DHT to data hiding is given.

## The Matrix Form of the DHT

The matrix form of the DHT requires that the data be of finite length. Since the DHT transform is defined for an infinite number of points, limitation of the DHT transform signal to a finite set would set up an approximation in the signal that is recovered.

The DHT is given below for data n=0, 1, 2, … :

$$
\begin{bmatrix} g(0) \\ g(1) \\ g(2) \\ g(3) \\ g(4) \\ g(5) \\ \cdot \\ \cdot \\ \cdot \end{bmatrix}
= \frac{2}{\pi}
\begin{bmatrix}
0 & \frac{1}{-1} & 0 & \frac{1}{-3} & 0 & \frac{1}{-5} & 0 & \frac{1}{-7} & \cdot \\
\frac{1}{1} & 0 & \frac{1}{-1} & 0 & \frac{1}{-3} & 0 & \cdot & \cdot & \cdot \\
0 & \frac{1}{1} & 0 & \frac{1}{-1} & 0 & \frac{1}{-3} & \cdot & \cdot & \\
\frac{1}{3} & 0 & \frac{1}{1} & 0 & \frac{1}{-1} & 0 & \cdot & \cdot & \cdot \\
0 & \frac{1}{3} & 0 & \frac{1}{1} & 0 & \frac{1}{-1} & \cdot & \cdot & \\
\frac{1}{5} & 0 & \frac{1}{3} & 0 & \frac{1}{1} & 0 & \cdot & \cdot & \\
0 & \cdot & \cdot & \cdot & \cdot & \cdot & \cdot & & \\
\frac{1}{7} & \cdot & & & & & & & \\
\cdot & \cdot & \cdot & \cdot & \cdot & \cdot & \cdot & \cdot &
\end{bmatrix}
\begin{bmatrix} f(0) \\ f(1) \\ f(2) \\ f(3) \\ f(4) \\ f(5) \\ \cdot \\ \cdot \\ \cdot \end{bmatrix}
$$

## DHT Based Randomness Measure

We define DHT based randomness measure as being equal to R. It is computed by first summing up the DHT values of the sequence and finding their average, which we represent by r. The randomness measure is 1 minus r.

$$r = \frac{1}{n} \sum_k g(k)$$

$$R = 1 - \left| \; r \; \right|$$

A variant of this measure will be to replace r by

$$r' = \left| \; \frac{1}{n} \sum_k \right| \; g(k) \; \|$$

giving

$$R' = 1 - r'$$

# Results and Discussions

We now provide results by considering various random switch sequences.

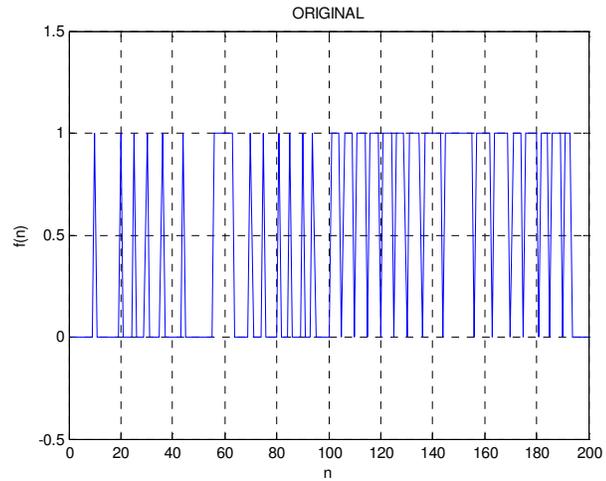

**Fig a**

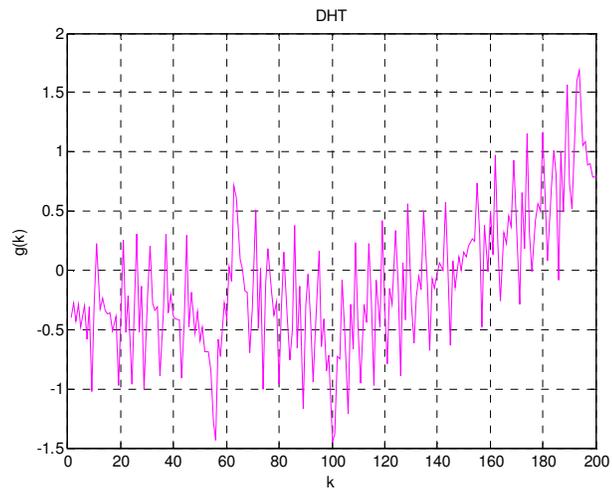

**Fig b**

Figure 3: Length: 200 - Switches: 13 (a) Original (b) DHT

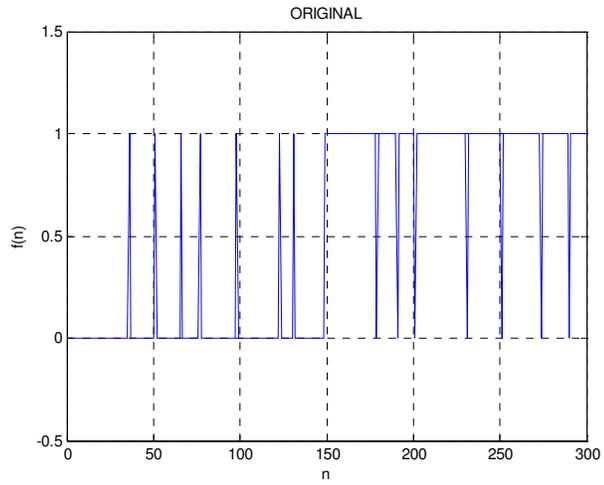

**Fig a**

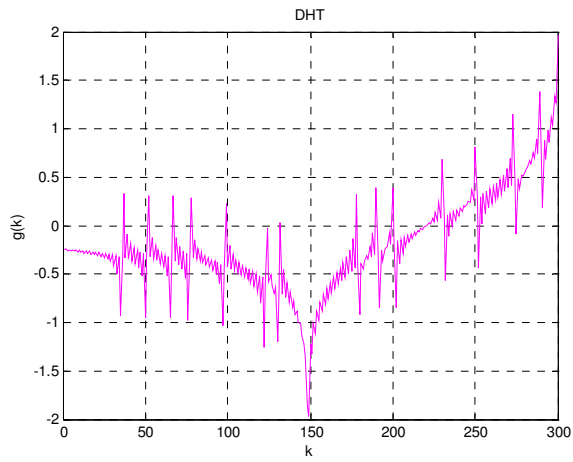

**Fig b**

Figure 4: Length: 300 - Switches: 7 (a) Original (b) DHT

We now consider prime reciprocal or D sequences.

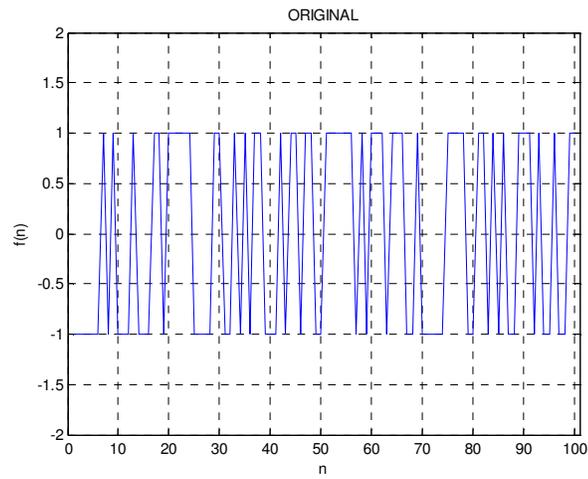

**Fig a**

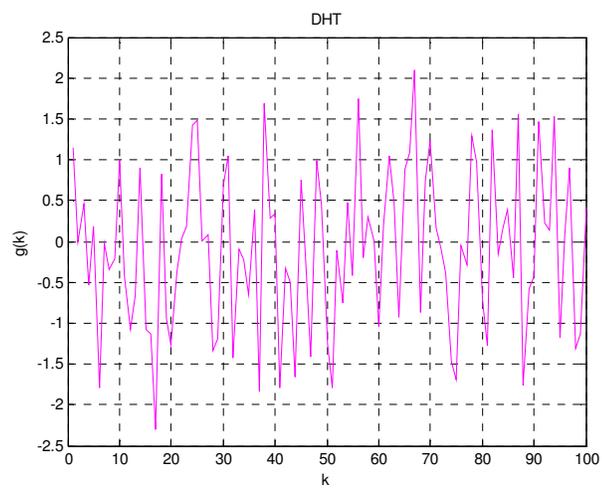

**Fig b**

Figure 5: 1/101 (a) Original (b) DHT

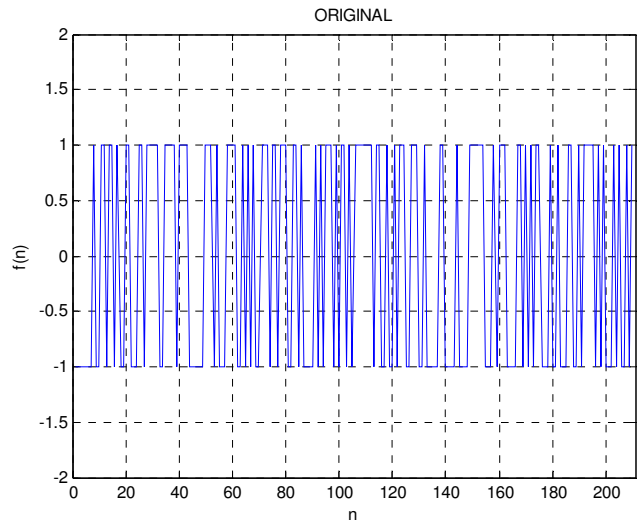

**Fig a**

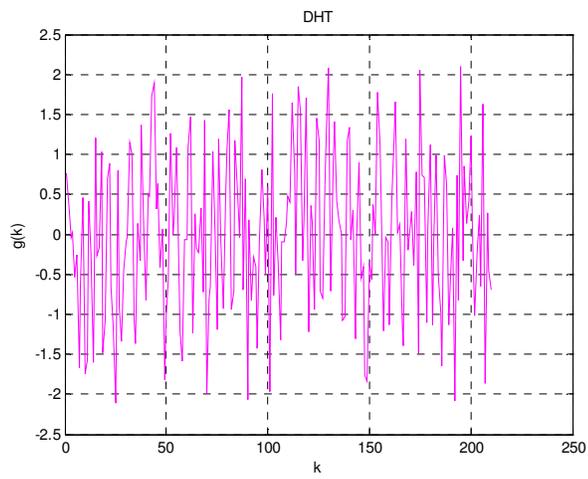

**Fig b**

Figure 6: 1/211 (a) Original (b) DHT

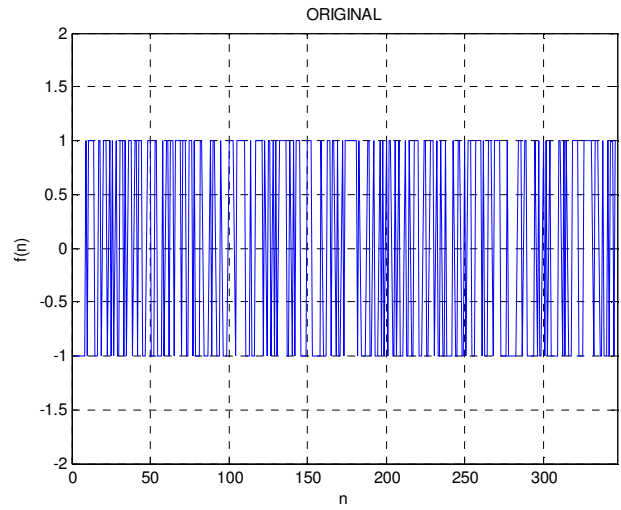

**Fig a**

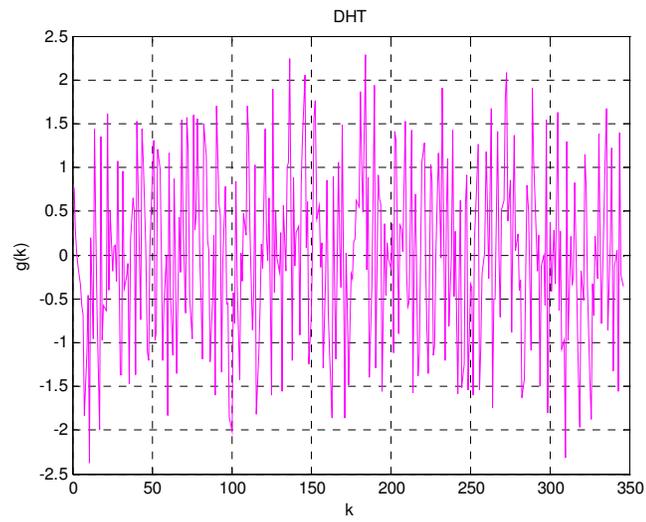

**Fig b**

Figure 7: 1/347 (a) Original (b) DHT

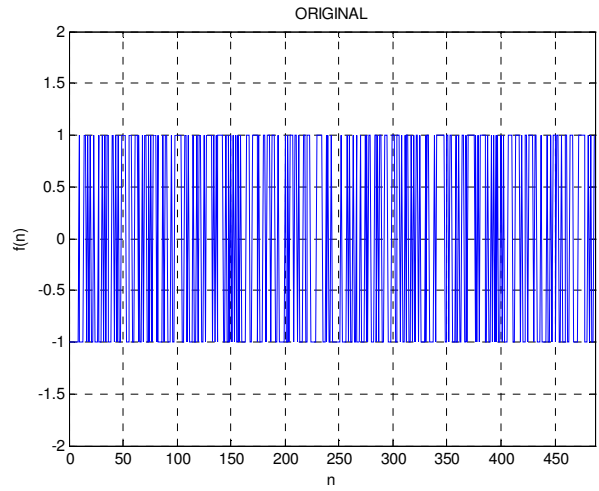

**Fig a**

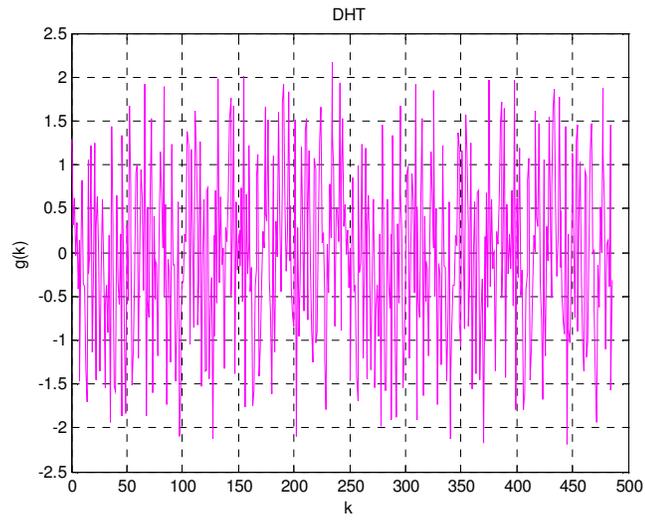

**Fig b**

Figure 8: 1/487 (a) Original (b) DHT

The randomness computed for different examples of random switch sequences are as follows:

| Length | No. of Switches | $R = 1 - r$ | $R' = 1 - r'$ |
|--------|-----------------|-------------|----------------|
| 100 | 1 | 0.7801 | 0.7783 |
| 100 | 3 | 0.8096 | 0.8043 |
| 100 | 4 | 0.8153 | 0.8150 |
| 100 | 5 | 0.8268 | 0.8233 |
| 100 | 7 | 0.8393 | 0.8310 |
| 100 | 10 | 0.8568 | 0.8526 |
| 100 | 13 | 0.8802 | 0.8793 |
| 100 | 20 | 0.8903 | 0.8803 |
| 200 | 4 | 0.7959 | 0.7832 |
| 200 | 5 | 0.7969 | 0.7910 |
| 200 | 7 | 0.7975 | 0.7934 |
| 200 | 11 | 0.7985 | 0.7964 |
| 200 | 13 | 0.8891 | 0.8810 |
| 200 | 20 | 0.9924 | 0.9915 |
| 300 | 4 | 0.7903 | 0.7893 |
| 300 | 5 | 0.7933 | 0.7912 |
| 300 | 7 | 0.7952 | 0.7946 |
| 300 | 11 | 0.7976 | 0.7965 |
| 300 | 13 | 0.8075 | 0.8036 |
| 300 | 20 | 0.9916 | 0.9825 |

As expected the R values increase with the number of switches and also with the increase in the length of the sequence.

The randomness computed for different examples of prime reciprocal or D-sequences are as follows:

| Prime Number | $R = 1 - r$ | $R' = 1 - r'$ |
|---|---|---|
| 1/13 | 0.7054 | 0.7036 |
| 1/67 | 0.8794 | 0.8723 |
| 1/127 | 0.9724 | 0.9690 |
| 1/151 | 0.9547 | 0.9710 |
| 1/223 | 0.9690 | 0.9758 |
| 1/331 | 0.9727 | 0.9765 |
| 1/463 | 0.9739 | 0.9790 |
| 1/557 | 0.9743 | 0.9810 |
| 1/631 | 0.9884 | 0.9845 |
| 1/821 | 0.9890 | 0.9867 |
| 1/991 | 0.9992 | 0.9943 |

We find that the results for D sequences are superior to those of switch sequences for small values of switches. But this changes as the number of switches becomes large. For a sequence of length 300, the switching sequence gives R value of 0.8075 and 0.9916 for 13 and 20 switches. Conversely, D sequence for 1/331 (comparable length) has a value of R that is 0.9727. Thus the D sequence is better than the comparable switch sequence with 13 switches but worse than a sequence with 20 switches.

The R values for random sequences generated by the computer are given next:

| Length | $R\ =\ 1-r$ | $R^{'} = 1 - r^{'}$ |
|---|---|---|
| 100 | 0.9653 | 0.9646 |
| 200 | 0.9656 | 0.9650 |
| 300 | 0.9767 | 0.9690 |
| 400 | 0.9779 | 0.9771 |
| 500 | 0.9782 | 0.9778 |
| 600 | 0.9889 | 0.9799 |
| 700 | 0.9893 | 0.9826 |
| 800 | 0,9998 | 0.9887 |

The randomness measure computed for different examples of computer generated random numbers superior to those obtained for prime reciprocal sequences.

## Conclusions

We find that the measure of randomness increases as the number of switches increases which we know from the idea of Kolmogorov computational complexity to have larger randomness [18]. We applied this measure to the switch sequences, prime reciprocal sequences, and to random numbers generated by the computer. Various experiments have been performed and the results were analyzed to verify the same. The results show that the proposed randomness measure does discriminate between sequences in a manner that is in accord with our expectations.

## Acknowledgements

I would also like to thank CTANS for supporting this research.